\documentclass[a4paper,fleqn]{cas-dc}
\usepackage[authoryear,longnamesfirst]{natbib}
\usepackage{subfig}
\usepackage{afterpage}
\usepackage{hyperref}
\hypersetup{
    colorlinks=true,
    linkcolor=blue,
    urlcolor=blue,
}
\def\tsc#1{\csdef{#1}{\textsc{\lowercase{#1}}\xspace}}
\tsc{WGM}
\tsc{QE}

\begin{document}
\let\WriteBookmarks\relax
\def\floatpagepagefraction{1}
\def\textpagefraction{.001}

\shorttitle{Deep Learning-based Earthquake Detection and Seismic Phase Classification}    
\shortauthors{Wei et al.}  

\title [mode = title]{Deep Learning-based Small Magnitude Earthquake Detection and Seismic Phase Classification}  

\author[1]{Wei Li}
\credit{Conceptualization, Methodology, Software, Writing - Original Draft, Writing - Review and Editing}
\author[2,1,3]{Yu Sha}
\credit{Methodology}

\author[1,4]{Kai Zhou}
\credit{Methodology, Writing - Review and Editing}

\author[1,4]{Johannes Faber}
\credit{Methodology, Writing - Review and Editing}

\author[1,5]{Georg Rümpker}
\credit{Conceptualization, Writing - Review and Editing}

\author[1,3,4,6]{Horst Stöcker}
\credit{Writing - Review and Editing}

\author[1,5]{Nishtha Srivastava}
\ead{srivastava@fias.uni-frankfurt.de}
\credit{Conceptualization, Methodology, Writing - Review and Editing}

\affiliation[1]{organization={Frankfurt Institute for Advanced Studies},
            city={Frankfurt am Main},
            postcode={60438}, 
            country={Germany}}
\affiliation[2]{organization={Key Laboratory of Intelligent Perception and Image Understanding of Ministry of Education, Academy of Advanced Interdisciplinary Research, Xidian University},
            city={Xian},
            postcode={710071}, 
            country={China}}
\affiliation[3]{organization={Xidian-FIAS international Joint Research Center, Giersch Science Center},
            city={Frankfurt am Main},
            postcode={60438}, 
            country={Germany}}
\affiliation[4]{organization={Institut f{\"u}r Theoretische Physik, Goethe Universit{\"a}t Frankfurt},
            city={Frankfurt am Main},
            postcode={60438}, 
            country={Germany}}
\affiliation[5]{organization={Institute of Geosciences, Goethe-University Frankfurt},
            city={Frankfurt am Main},
            postcode={60438}, 
            country={Germany}}
\affiliation[6]{organization={GSI Helmholtzzentrum f{\"u}r Schwerionenforschung GmbH},
            city={Darmstadt},
            postcode={64291}, 
            country={Germany}}

\cortext[1]{Corresponding author}



\begin{abstract}
Reliable earthquake detection and seismic phase classification is often challenging especially in the circumstances of low magnitude events or poor signal-to-noise ratio. With improved seismometers and better global coverage, a sharp increase in the volume of recorded seismic data is witnessed. This makes the handling of the seismic data rather daunting based on traditional approaches and therefore fuels the need for a more robust and reliable method. In this study, we investigate two deep learning-based models, termed 1D Residual Neural Network (ResNet) and multi-branch ResNet, for tackling the problem of seismic signal detection and phase identification, especially the later can be used in the case where multiple classes is organized in the hierarchical format. These methods are trained and tested on the dataset of the Southern California Seismic Network. Results demonstrate that the proposed methods can achieve robust performance for the detection of seismic signals, and the identification of seismic phases, even when the seismic events are of small magnitude and are masked by noise. Compared with previously proposed deep learning methods, the introduced frameworks achieve 4$\%$ improvement in earthquake monitoring, and a slight enhancement in seismic phase classification.
\end{abstract}

% \begin{highlights}
% \item 1D ResNet34 architecture is proposed and applied to detect earthquake signals and identify seismic phases.
% \item 1D multi-branch ResNet model is developed to simultaneously complete the task of earthquake detection and seismic phase classification.
% \item Compared with the baseline methods, the proposed methods achieve superior results.
% \end{highlights}

% Keywords
% Each keyword is separated by \sep

\maketitle

% Main text
\section{Introduction}\label{Introduction}
The detection of earthquake events is crucial for seismologists to optimally monitor tectonic activities in a region. In order to achieve reliable earthquake monitoring, many automated methods for seismic phase picking have been developed. The most state-of-the-art conventional algorithms for earthquake detection or seismic phase picking include template matching \cite{peng2009migration, ross2017aftershocks}, short-time average/long-time average (STA/LTA) \cite{allen1978automatic}, and autoregressive Akaike Information Criterion-picker (AR-AIC) \cite{sleeman1999robust}. Template matching involves measuring the similarity between earthquake waveforms and catalogued waveforms of seismic events. STA/LTA refers to measuring the ratio between the amplitude of the signal on a short time window with that on a long time window, and the detection can be determined when the ratio is greater than some pre-defined threshold. AR-AIC involves two main calculations: AR and and AIC, where the minimum of the two-model AIC is identified as the phase arrival time, and this method is mainly based on the assumption that the seismogram can be divided into different locally stationery segments \cite{st2011akaike}. However, template matching relies heavily on the pre-defined events, this makes the detection challenging in case of unfamiliar data \cite{ross2018generalized}. On the other hand, both STA/LTA and AIC do not perform well for the signals with low signal-to-ratio (SNR) signal especially in the case of low magnitude events. In addition, given seismic recordings associated with low SNRs, AR-AIC is potential to produce several local minimal AIC value that results in false arrival time picking \cite{dong2019arrival}.

Over the past decades, due to the development of seismic equipment and seismic monitoring network, remarkable improvements have been achieved in seismic event detection system which brings about huge and rapidly-increasing seismic database. This thus calls for robust and sensitive methods to address the ever growing volume of seismic data.  Therefore, seismic event detection and phase picking algorithms are becoming increasingly important to automatically deal with large seismic data. Deep learning, with its recent development - especially in computer vision, is capable to process big data and with a large number of different features such as lines, edges, image segments. The task of earthquake signal detection or seismic phase identification can be recognized as similar to the identification of objects in computer vision. Therefore, the recent advances in the field of computer vision have great potential in seismological applications. Recently, deep learning has been widely used to detect earthquakes or identify seismic phases \cite{ross2018generalized, saad2020earthquake, saad2021capsphase, chakraborty2021study}. For example, in \cite{ross2018generalized}, a convolutional neural network was trained on the huge amount of labeled seismic data to classify seismic body wave phase. 
% This creates a need for efficient algorithms for reliable detection.

In this work, both the earthquake detection and seismic phase identification are formulated as a supervised classification problem. Considering that ResNet \cite{he2016deep} achieves superiority in image classification by adopting skip connection and 1D ResNet works well for time-series data (e.g., valve acoustic signals) \cite{sha2022multi}, in this study we investigate two deep learning-based methods: 1D ResNet34 based on the residual module \cite{he2016deep} and 1D multi-branch ResNet for the defined task (more details in section 2). The seismic data of the Southern California Seismic Network \cite{scec2013} labeled as 'P-phase', 'S-phase' and 'Noise' (more details in section 3.1) is employed to train the model and test it's performance.  The model performance well indicates that the proposed methods for earthquake detection are not only capable of robustly classifying seismic signals and noise data, but also allows to reliably identify P-waves and S-waves of small magnitude earthquakes.

This work is organized as follows. Section 2 comprehensively delineates the proposed methodologies. Section 3 briefly describes the used dataset, and shows the experiment setting and metrics for performance evaluation. Section 4 details the results of the performed experiments. Finally, Section 5 describes the conclusions of this work.

\section{Methodology}
The proposed methods described in this section are taking a window of three-channel waveform seismogram data (e.g., three-channel normalized waveform within the duration of 4s) as input. Note that in this work the two tasks (earthquake detection and Phase classification) are separately implemented when using 1D ResNet, while they are completed simultaneously in the case of multi-branch ResNet with reduced number of model parameters. Therefore, in the task of earthquake detection, the output is labeled as earthquake (including but not distinguishing P wave and S wave) and noise signals same as \cite{saad2020earthquake}, while in the case of seismic phase classification, the model is trained to classify noise, P wave, and S wave, respectively, similar to \cite{ross2018generalized,saad2021capsphase}. 

For those two tasks, in the model training process, the labels are defined as follows: (i) for earthquake detection - 'zero' for earthquake signal and 'one' for noise. (ii) for phase classification - 'zero' for P-wave windows, 'one' for S-wave windows, and 'two' for noise windows.

Considering the complexity during training and memory consumption, in the multi-branch architecture, two branches are used with one coarse branch for earthquake and noise identification, and one fine branch for seismic phase classification (P-phase, S-phase, Noise). 

\subsection{Residual neural network}
\begin{figure}[t]
    \centering
    \subfloat{\includegraphics[width=0.5\textwidth]{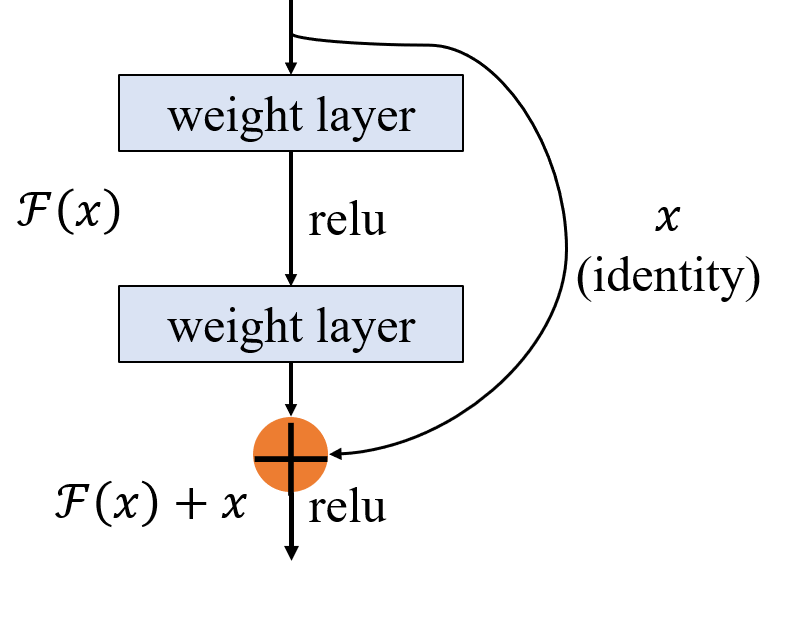}}
    \caption{Residual block \cite{he2016deep}. $x$ is the input vector of the layer, and $\mathcal{F}(x)$ denotes the residual function to be learned.}
    \label{fig1}
\end{figure}

 \cite{he2016deep} revealed that when adding more layers to the neural network to enrich the features of the model, training the neural network becomes more challenging due to difficulty in optimizing the model parameters caused by vanishing/exploding gradient. Furthermore, the accuracy of the model either gets saturated at a particular value, or slowly degrade. Consequently, the model performance deteriorates both in the training phase and testing phase. To tackle this issue, He \textit{et al.} \cite{he2016deep} proposed the popular neural network architecture known as ResNet, where the layers are explicitly reformulated as learning residual functions with respect to the layer input. The idea is that it is easier to optimize the residual function $\mathcal{F}(x)$ than the original function $H(x) = \mathcal{F}(x)+x$. They also provided extensive empirical evidence to indicate that it is not only easier to optimize these residual networks, but also the accuracy could be enhanced from the considerably increased depth. The Residual block is denoted in Figure \ref{fig1}. 
% due to difficulty in optimizing the model parameters and vanishing gradients. Gradients are used to update the weights, however, if the network is too deep gradient becomes small, this further prevents the weights to be updated

The skip connections in ResNet succeeded in dealing with the issue of vanishing gradient in deep neural networks by allowing the gradient to flow directly through the alternative shortcut path backward from latter layers to former layers. On the other hand, these connections allow the model to learn the identity functions, which guarantees that the higher layer could perform at least as good as the lower layer, and not worse.

In this study, 1D ResNet is developed for earthquake detection and seismic phase classification. The only difference between these two tasks is the output size. In our work, the seismic records with three-components are identified as images by the devised 1D ResNet to detect earthquake signals from noisy data and further to identify seismic phases to P-wave or S-wave, correspondingly.

\subsection{ Multi-branch Residual Neural Network}
\begin{figure*}[t]
    \centering
    \subfloat{\includegraphics[width=\textwidth]{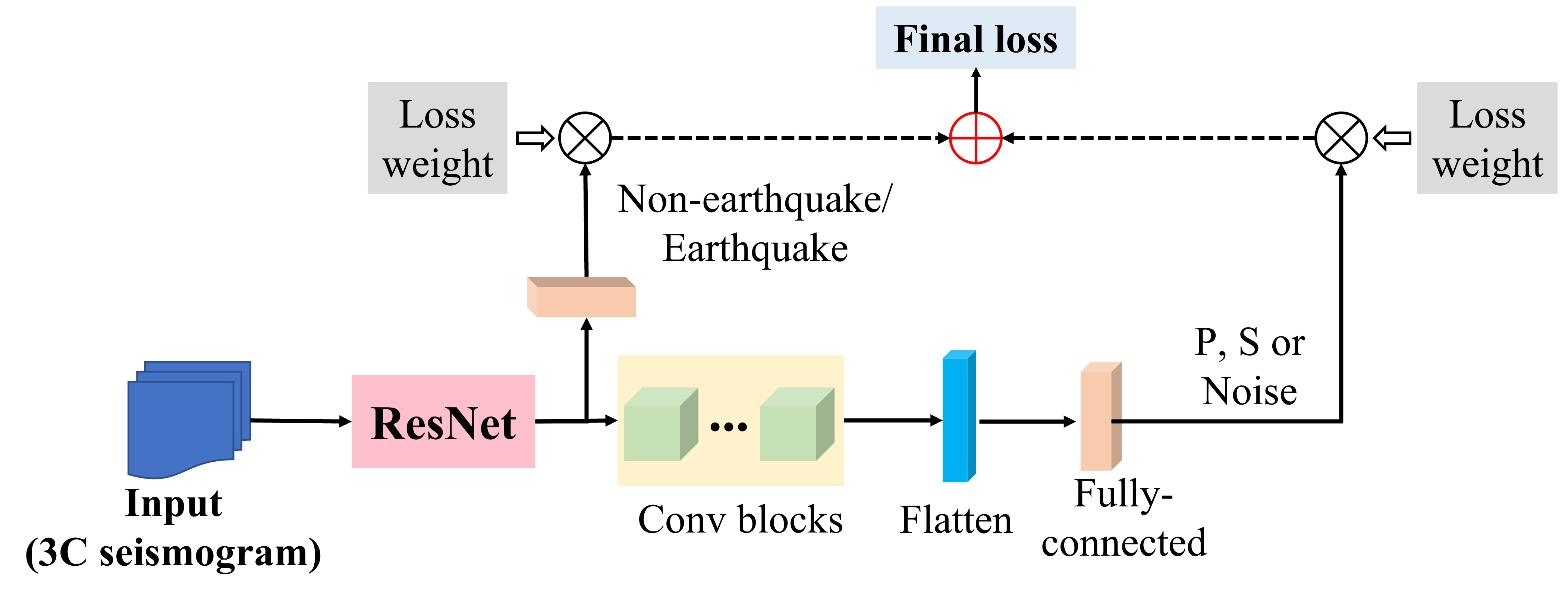}}
    \caption{Architecture of Multi-Branch ResNet. The network at the bottom can be an arbitrary ResNet. There can be multiple branch networks and each of them outputs a prediction. The final loss function is a weighted sum over all branch losses.}
    \label{fig2}
\end{figure*}

 A 1D multi-branch ResNet based architecture is developed to perform earthquake detection and seismic phase classification simultaneously. This model combines several branch networks and the main ResNet workflow to perform classifications hierarchically. Here, the prior knowledge of class hierarchical structure is used to improve the model's learning ability. The architecture is shown in Figure \ref{fig2}.

\begin{figure}[t]
    \centering
    \subfloat{\includegraphics[width=0.5\textwidth]{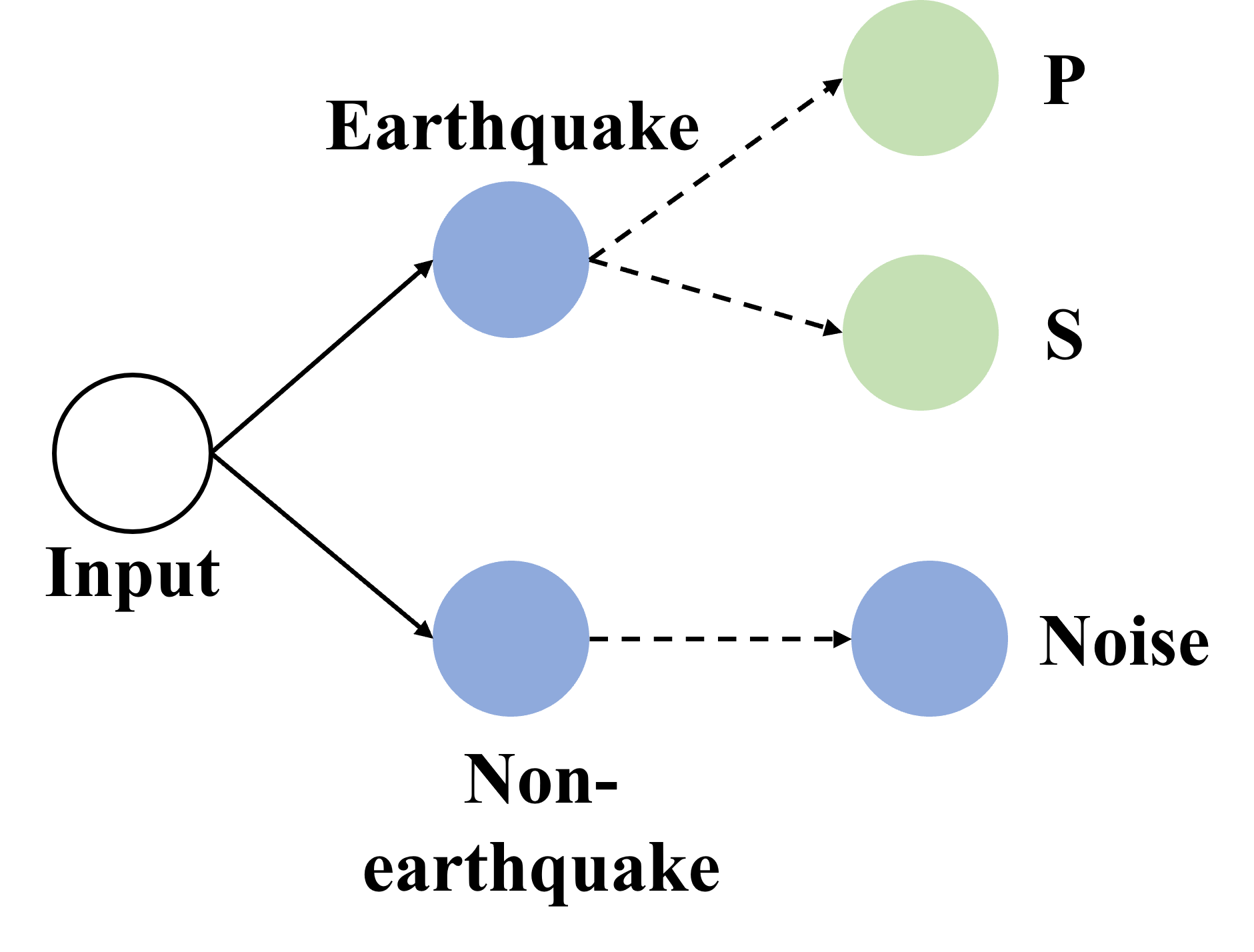}}
    \caption{ A sample hierarchical label tree where classes are taken from SCEDC dataset.}
    \label{fig3}
\end{figure}

In the multi-branch ResNet model, the existent ResNet component is utilized as the fundamental block to build a neural network combined with different branches. The network displayed at the bottom in Figure \ref{fig2} is comprised of a random convolutional block with multiple layers. The output branches of the multi-branch ResNet are shown in the upper part of Figure \ref{fig2}. As presented in Figure \ref{fig2}, each branch generates a prediction corresponding to the level in the label tree described as Figure \ref{fig3}. For each branch network, fully connected layers along with a softmax layer are leveraged to get the probability for each class, and then the final predicted label can be formatted in a one-hot vector. 

In this study, the loss function of the developed 1D multi-branch ResNet can be achieved by summing all branch losses assigned with different weights. Hence, the loss function can defined as follows:
\begin{equation}
    \mathcal{L}_{total} = w_1* loss_1 + w_2*loss_2
\end{equation}
where $w_1$ and $w_2$ are the weights for the noise/earthquake detection, and seismic phase classification, respectively. And $loss_1$ and $loss_2$ denote the losses of two branches. In this paper, we set $w_{branch 1} = w_{branch 2} = 0.5$ for simplicity. 

\section{Analysis}
\subsection{Seismic Dataset}
\begin{figure*}[t]
    \centering
    \subfloat{\includegraphics[width=\textwidth]{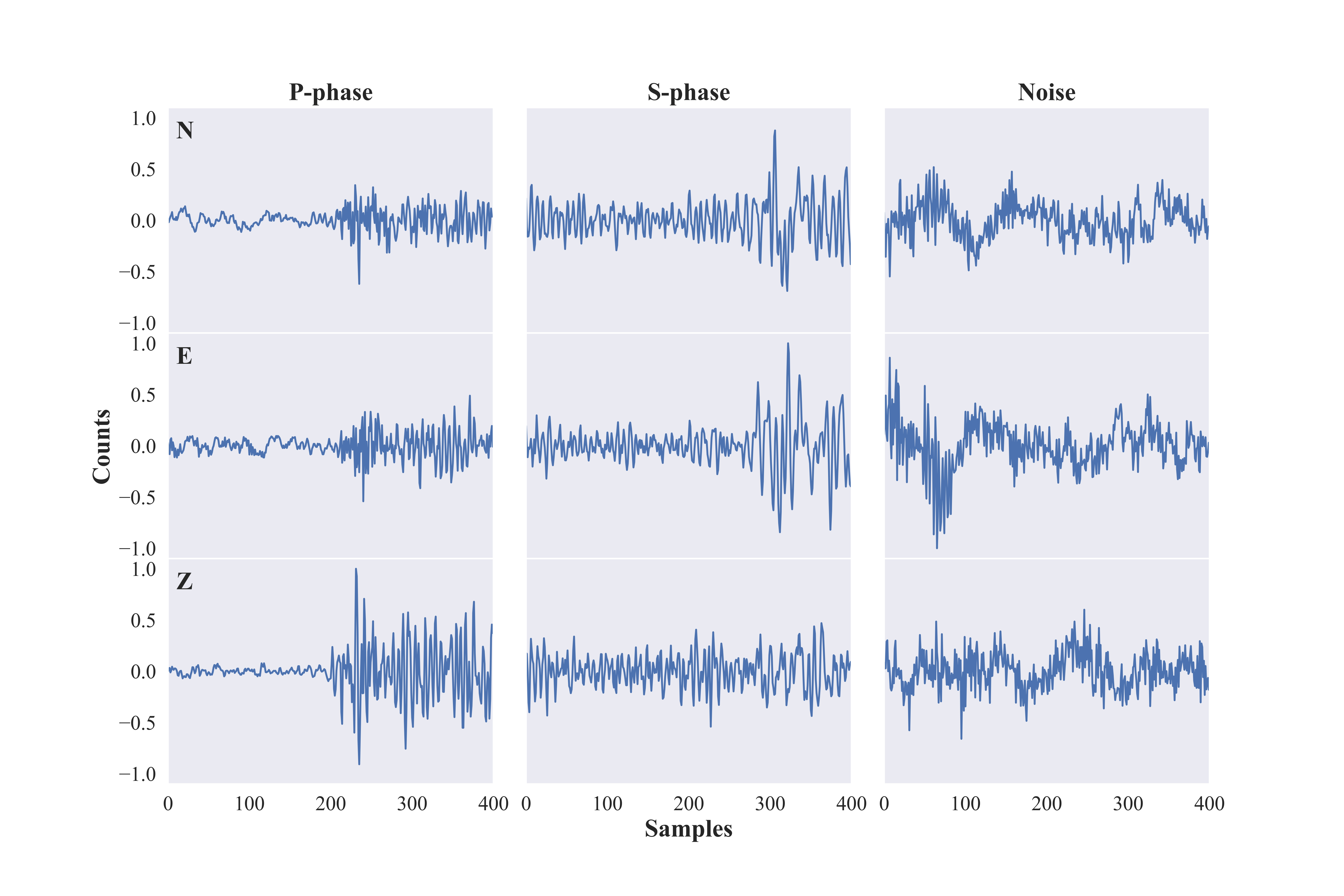}}
    \caption{Waveform visualization of SCEDC data \cite{scec2013}. These waveform is resampled at 100Hz and normalized using the absolute maximum amplitude over three components.}
    \label{fig4}
\end{figure*}

Within this study, the dataset provided by Southern California Earthquake Data Center (SCEC) \cite{scec2013} is utilized to train and test the proposed model. The magnitude range of the data is $-0.81<M<5.7$. The dataset includes 1.5 million P-wave seismograms, 1.5 million S-wave seismograms, and 1.5 million noise windows with each record of 4s duration. Both P-wave and S-wave windows are centered on the arrival pick, while each noise window is captured by starting 5s before each P-wave arrival pick. Figure \ref{fig4} visualizes the waveforms including P-phase window, S-phase window and noise window.

\begin{figure*}[t]
    \centering
    \subfloat{\includegraphics[width=\textwidth]{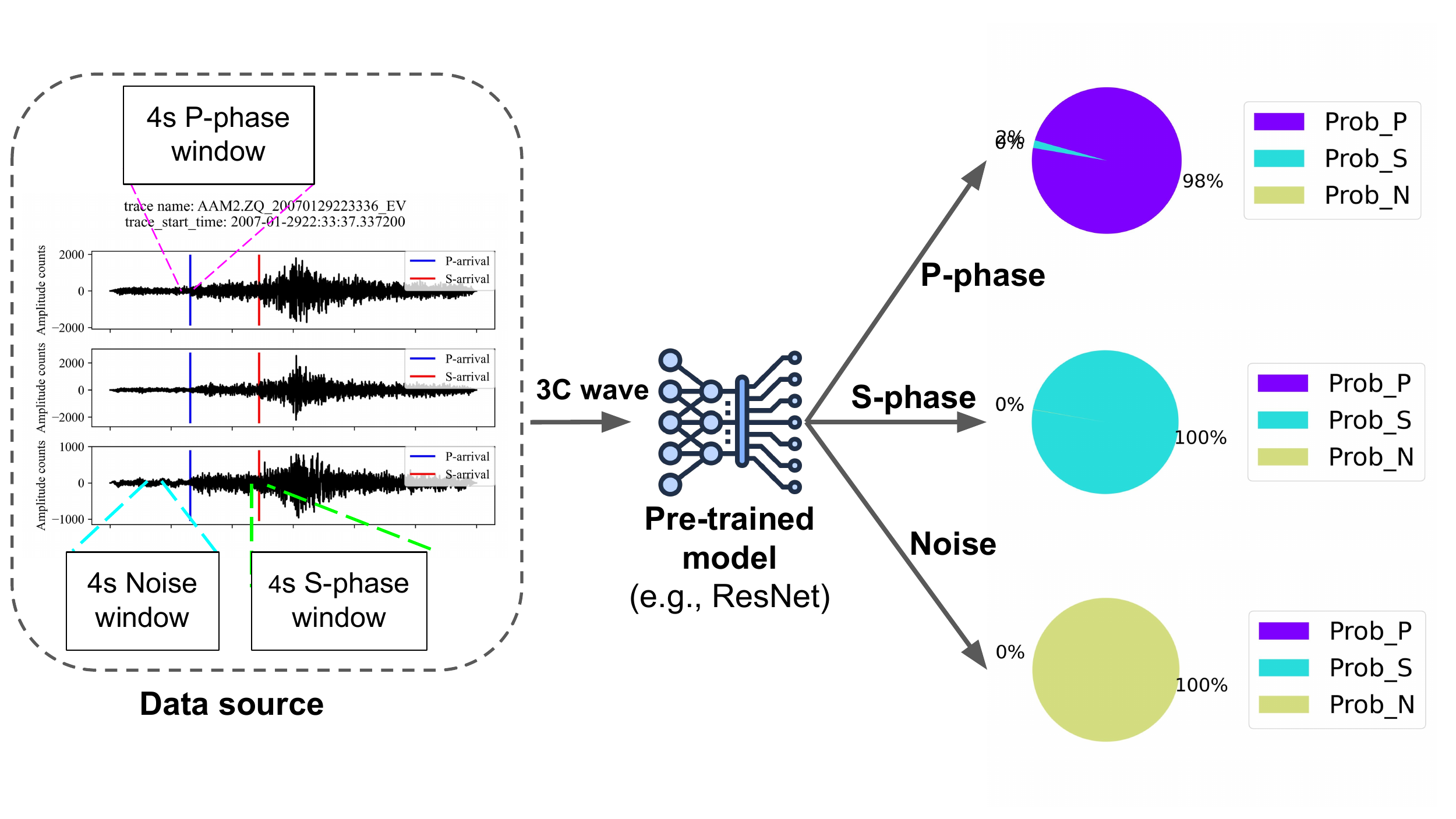}}
    \caption{Output visualization of one testing example from STEAD dataset \cite{mousavi2019stanford}, where the pie chart displays the predicted probability for the classes corresponding to different input data.}
    \label{fig5}
\end{figure*}

In order to help better understand the model output, here Figure \ref{fig5} shows the pipeline of the testing process, and gives a visualization of the testing result (the predicted probability) using the pie chart of three different input data extracted from one raw waveform of STEAD dataset \cite{mousavi2019stanford} when fed in the pre-trained model. Similar to SCEDC dataset \cite{scec2013}, the P-phase and S-phase windows are centered on the respectively arrival time, and the noise windows are extracted starting 5s before the P arrival time. Each pie chart, shows the predicted probability for each class (P-phase, S-phase, Noise), in which the sum over all the probability is one.

\begin{table*}[t]
    \caption{Testing accuracy for two tasks.}
    \label{table1}
    \centering
    \subfloat{
        \begin{tabular}{ccc}
        \hline
        Model & \begin{tabular}[c]{@{}c@{}}Earthquake detection\end{tabular} & \begin{tabular}[c]{@{}c@{}}Phase classification\end{tabular} \\
        \hline
        CapsNet \cite{saad2020earthquake} (50\% training)      & 98.40\%  & -        \\
        CapsPhase \cite{saad2021capsphase} (90\% training)      & -        & 98.67\%  \\
        1D ResNet                                  & \textbf{98.83\%}  & \textbf{98.70\%}  \\
        Multi-branch ResNet (50\% training)    &\textbf{98.85\%}  & 98.41\%  \\
        Multi-branch ResNet (90\% training)     & 98.96\%  & \textbf{98.66\%}  \\
        \hline
        \end{tabular}
    }
\end{table*}

\subsection{Experiment Setting}
In this study, the learning rate of $0.001$ is used and the model is trained till 50 epochs same as CapsPhase \cite{saad2021capsphase} to achieve an unbiased comparison. The proposed models are implemented in Pytorch \cite{paszke2019pytorch} and trained on an NVIDIA A100 Graphics Processing Unit. The ADAM \cite{kingma2014adam} algorithm is adapted to optimize our models by using a cross-entropy loss function in the mini-batches of 480 records. A dropout rate of 0.2 is adopted for all dropout layers. Note that here, the data augmentation is not used on the training data and at the same time, we do not utilize any ensemble methods in the model training phase. 

Aiming to make the comparison with CapsNet \cite{saad2020earthquake} for the earthquake detection task, 50\% of the whole dataset is used for model training and 25\% of the data is utilized for testing. However, in CapsNet \cite{saad2020earthquake}, the training dataset is balanced, i.e., the number of the data labeled by zero ('earthquake') is same as the number of the data with labels of one ('noise'). Please note the fact that, from the beginning on, the whole dataset is labelled with one of the three classes: P-phase, S-phase or noise. In our study, the first 50$\%$ of the whole data is used for training. On other hand, in the case of earthquake and non-earthquake detection, the wave windows including P-phase and S-phase are re-labeled as earthquake signals. The above-mentioned processes make the training dataset biased, where the number of the data labeled by zero (earthquake) is not equal to that labeled by one (noise). Owing to this, it is more difficult to train the model in our work. This can further test the model performance like robustness on imbalanced data which is common in real-world applications of deep learning. Furthermore, for seismic phase identification, we split the seismograms into a training set (90$\%$) and testing set (5$\%$) same as CapsPhase \cite{saad2021capsphase}.  Nevertheless, it is worth noting that only in the task for earthquake identification, the training dataset is imbalanced, while in the seismic phase classification, the training dataset is unbiased.

\subsection{Evaluation Metric}
The following metrics are utilized to evaluate the model performance. First, the accuracy is defined as the ratio of correctly identified instances over all testing samples, which is usually regarded as the basic measurement of a classifier’s performance.
\begin{equation}
    Accuracy = \frac{N_C}{N_T}
\end{equation}
where $N_C$ denotes the number of correctly labeled samples and $N_T$ represents the number of all testing samples.

Then, in order to further estimate the model’s effectiveness, the confusion matrix \cite{stehman1997selecting} is employed to reflect the classification result. Furthermore, given a confusion matrix, the precision, recall and F1-score can be defined as follows:
\begin{equation}
    Precision = \frac{TP}{TP+FP}
\end{equation}
\begin{equation}
    Recall = \frac{TP}{TP+FN}
\end{equation}
\begin{equation}
    F1_{scores} = 2*\frac{Precision * Recall}{Precision + Recall}
\end{equation}
where TN, FN, FP, and TP are the true negative, false negative, false positive, and true positive, respectively.

\section{Discussions}
The overall testing accuracy for earthquake detection and phase classification of different methods is compared and summarized in Table \ref{table1}. We can find that the testing accuracy of 1D ResNet and multi-branch ResNet is 98.83$\%$ and 98.85$\%$, respectively. The result demonstrates that our proposed model achieves better performance for earthquake detection over CapsNet \cite{saad2020earthquake}. For seismic phase classification, 1D ResNet demonstrates its superiority, and multi-branch ResNet also achieves a compatible performance compared with CapsPhase \cite{saad2021capsphase}. The potential reasons to achieve higher performance are twofold. First, the residual blocks in ResNet \cite{he2016deep} contribute to improve the classification accuracy, since it is capable to learn some meaningful features from the input. Second, the hierarchical structure in multi-branch ResNet as prior knowledge could have enhanced model performance.

\begin{table*}[t]
    \caption{Testing results for earthquake detection on the testing data.}
    \label{table2}
    \centering
    \begin{tabular}{c|cccc}
    \hline
                Category             & Model        & Precision & Recall & F1-score \\
    \hline
    \multirow{3}{*}{Earthquake } & CapsNet \cite{saad2020earthquake} & 98.64\%  & 98.98\%  & 98.80\%  \\
                             &  ResNet                            & \textbf{99.18\%}  & 98.06\%  & \textbf{99.12\%}  \\
                             &  Multi-branch ResNet (50\% training)         & 97.77\%  & 98.82\%  & 98.29\%  \\
                        
    \hline
    \multirow{3}{*}{Noise} & CapsNet \cite{saad2020earthquake}       & 97.96\%  & 97.30\%  & 98.70\%   \\
                             &  ResNet                            & 98.13\% & 98.37\%  & 98.25\%   \\
                             & Multi-branch ResBet (50\% training)         & \textbf{99.40\%} & \textbf{98.87\%} & \textbf{99.14\%}   \\
    \hline
    \end{tabular}
\end{table*}
\begin{table*}[t]
    \centering
    \caption{Testing results for phase classification on the testing data.}
    \label{table3}
    \begin{tabular}{c|cccc}
    \hline
              Category               & Model        & Precision & Recall & F1-score \\
    \hline
    \multirow{3}{*}{P phase} & CapsPhase \cite{saad2021capsphase}    & 98.68\%  & 98.99\%  & 98.76\%  \\
                             &  ResNet                            & \textbf{98.88\%} & 98.64\%  & \textbf{98.76\%}  \\
                             &Multi-branch ResNet (90\% training)         & \textbf{98.84\%}  & 98.58\%  & 98.71\%  \\
    \hline
    \multirow{3}{*}{S Phase} & CapsPhase \cite{saad2021capsphase}    & 98.40\%  & 98.88\%  & 98.70\%  \\
                             &  ResNet                            & \textbf{98.72\%}  & \textbf{98.94\%}  & \textbf{98.83\%}  \\
                             &Multi-branch ResBet (90\% training)         & \textbf{98.88\%}  & 98.68\%  & \textbf{98.78\%}  \\
    \hline
    \multirow{3}{*}{Noise}   & CapsPhase \cite{saad2021capsphase}    & 98.90\%     & 98.17\%       & 98.53\%        \\
                             &  ResNet                            & 98.52\%  & \textbf{98.54\%}  & \textbf{98.53\%}  \\
                             &Multi-branch ResBet (90\% training)         & 98.26\%  & \textbf{98.73\%} & 98.49\%  \\
    \hline
    \end{tabular}
\end{table*}
\begin{figure*}[t]
    \centering
    \includegraphics[width=\textwidth]{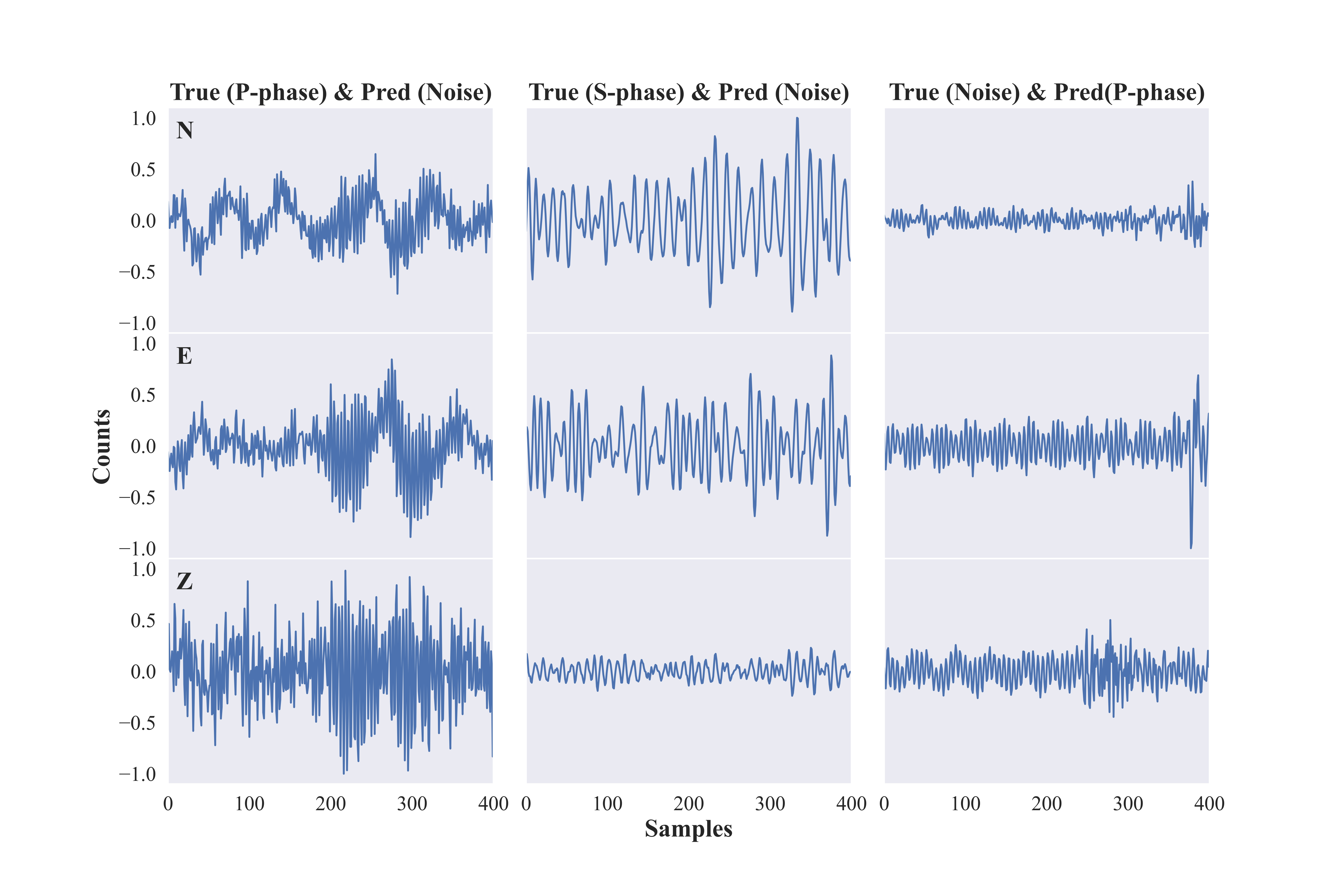}
    \caption{Visualization of mis-classified SCEDC data. These waveforms are resampled at 100Hz and normalized using the absolute maximum amplitude over three components.}
    \label{fig6}
\end{figure*}

\cleardoublepage 
\cleardoublepage 
The results of different metrics including precision, recall and F1-score for earthquake identification are shown in Table \ref{table2}. It is found that 1D ResNet and multi-branch ResNet can achieve compatible results. Particularly, for noise detection, the proposed multi-branch ResNet reaches a best performance. 

Finally, Table \ref{table3} summarizes the classification result of different metric for seismic phase identification. It can be observed that compared with CapsPhase \cite{saad2021capsphase}, the proposed models achieve a better performance, especially for P wave and S wave identification.

Figure \ref{fig6} visualizes an example for each class including P-phase window, S-phase window and noise window from SCEDC dataset that are mis-classified using the proposed 1D-ResNet in this work. 

\section{Conclusions}
In this study, we investigate two deep learning-based models for simultaneous earthquake detection and seismic phase classification. The methods are built based on a well-designed architecture using a 1D residual neural network (ResNet). These models are trained and tested on the Southern California Seismic Dataset. Extensive experiment results verify that both the used method and the proposed multi-branch ResNet achieve better performance than previous deep learning based approaches. The proposed model can be utilized by seismologist to identify the earthquake signals and phase identification, especially in the case of noisy low magnitude earthquake waveforms. The future work will focus on a straightforward hierarchical classifier to reduce the training complexity and memory.

\section{Code and Data Availability}
The code is available on GitHub at \url{https://github.com/srivastavaresearchgroup/Seismic-phase-Classification}. The Southern California seismic data that support this study can be accessed in \cite{scec2013}.

\section*{Acknowledgment}
This work is supported by supported by the ''KI-Nachwu
chswissenschaftlerinnen" - grant SAI 01IS20059 by the Bundesministerium für Bildung und Forschung - BMBF. Calculations were performed at the Frankfurt Institute for Advanced Studies' new GPU cluster, funded by BMBF for the project Seismologie und Artifizielle Intelligenz (SAI). Horst Stöcker gratefully acknowledges the Judah M. Eisenberg Laureatus - Professur at Fachbereich Physik, Goethe Universität Frankfurt, funded by the Walter Greiner Gesellschaft zur Förderung der physikalischen Grundlagenforschung e.V.
\printcredits
\bibliographystyle{cas-model2-names}

\bibliography{cas-refs}

\begin{thebibliography}{17}
\expandafter\ifx\csname natexlab\endcsname\relax\def\natexlab#1{#1}\fi
\providecommand{\url}[1]{\texttt{#1}}
\providecommand{\href}[2]{#2}
\providecommand{\path}[1]{#1}
\providecommand{\DOIprefix}{doi:}
\providecommand{\ArXivprefix}{arXiv:}
\providecommand{\URLprefix}{URL: }
\providecommand{\Pubmedprefix}{pmid:}
\providecommand{\doi}[1]{\href{http://dx.doi.org/#1}{\path{#1}}}
\providecommand{\Pubmed}[1]{\href{pmid:#1}{\path{#1}}}
\providecommand{\bibinfo}[2]{#2}
\ifx\xfnm\relax \def\xfnm[#1]{\unskip,\space#1}\fi
%Type = Article
\bibitem[{Allen(1978)}]{allen1978automatic}
\bibinfo{author}{Allen, R.V.}, \bibinfo{year}{1978}.
\newblock \bibinfo{title}{Automatic earthquake recognition and timing from
  single traces}.
\newblock \bibinfo{journal}{Bulletin of the seismological society of America}
  \bibinfo{volume}{68}, \bibinfo{pages}{1521--1532}.
%Type = Article
\bibitem[{Center(2013)}]{scec2013}
\bibinfo{author}{Center, S.C.E.D.}, \bibinfo{year}{2013}.
\newblock \bibinfo{title}{Southern california earthquake data center (2013)}.
\newblock \bibinfo{journal}{California Institute of Technology, Dataset}
  \bibinfo{note}{Doi:{10.7909/C3WD3xH1}}.
%Type = Article
\bibitem[{Chakraborty et~al.(2021)Chakraborty, Li, Faber, Ruempker, Stoecker
  and Srivastava}]{chakraborty2021study}
\bibinfo{author}{Chakraborty, M.}, \bibinfo{author}{Li, W.},
  \bibinfo{author}{Faber, J.}, \bibinfo{author}{Ruempker, G.},
  \bibinfo{author}{Stoecker, H.}, \bibinfo{author}{Srivastava, N.},
  \bibinfo{year}{2021}.
\newblock \bibinfo{title}{A study on the effect of input data length on deep
  learning based magnitude classifier}.
\newblock \bibinfo{journal}{arXiv preprint arXiv:2112.07551} .
%Type = Article
\bibitem[{Dong et~al.(2019)Dong, Jiang, Li and Yang}]{dong2019arrival}
\bibinfo{author}{Dong, X.}, \bibinfo{author}{Jiang, H.}, \bibinfo{author}{Li,
  Y.}, \bibinfo{author}{Yang, B.}, \bibinfo{year}{2019}.
\newblock \bibinfo{title}{Arrival time picking of microseismic data by using
  spe algorithm}.
\newblock \bibinfo{journal}{JOURNAL OF SEISMIC EXPLORATION}
  \bibinfo{volume}{28}, \bibinfo{pages}{475--494}.
%Type = Inproceedings
\bibitem[{He et~al.(2016)He, Zhang, Ren and Sun}]{he2016deep}
\bibinfo{author}{He, K.}, \bibinfo{author}{Zhang, X.}, \bibinfo{author}{Ren,
  S.}, \bibinfo{author}{Sun, J.}, \bibinfo{year}{2016}.
\newblock \bibinfo{title}{Deep residual learning for image recognition}, in:
  \bibinfo{booktitle}{Proceedings of the IEEE conference on computer vision and
  pattern recognition}, pp. \bibinfo{pages}{770--778}.
%Type = Article
\bibitem[{Kingma and Ba(2014)}]{kingma2014adam}
\bibinfo{author}{Kingma, D.P.}, \bibinfo{author}{Ba, J.}, \bibinfo{year}{2014}.
\newblock \bibinfo{title}{Adam: A method for stochastic optimization}.
\newblock \bibinfo{journal}{arXiv preprint arXiv:1412.6980} .
%Type = Article
\bibitem[{Mousavi et~al.(2019)Mousavi, Sheng, Zhu and
  Beroza}]{mousavi2019stanford}
\bibinfo{author}{Mousavi, S.M.}, \bibinfo{author}{Sheng, Y.},
  \bibinfo{author}{Zhu, W.}, \bibinfo{author}{Beroza, G.C.},
  \bibinfo{year}{2019}.
\newblock \bibinfo{title}{Stanford earthquake dataset (stead): A global data
  set of seismic signals for ai}.
\newblock \bibinfo{journal}{IEEE Access} \bibinfo{volume}{7},
  \bibinfo{pages}{179464--179476}.
%Type = Article
\bibitem[{Paszke et~al.(2019)Paszke, Gross, Massa, Lerer, Bradbury, Chanan,
  Killeen, Lin, Gimelshein, Antiga et~al.}]{paszke2019pytorch}
\bibinfo{author}{Paszke, A.}, \bibinfo{author}{Gross, S.},
  \bibinfo{author}{Massa, F.}, \bibinfo{author}{Lerer, A.},
  \bibinfo{author}{Bradbury, J.}, \bibinfo{author}{Chanan, G.},
  \bibinfo{author}{Killeen, T.}, \bibinfo{author}{Lin, Z.},
  \bibinfo{author}{Gimelshein, N.}, \bibinfo{author}{Antiga, L.}, et~al.,
  \bibinfo{year}{2019}.
\newblock \bibinfo{title}{Pytorch: An imperative style, high-performance deep
  learning library}.
\newblock \bibinfo{journal}{arXiv preprint arXiv:1912.01703} .
%Type = Article
\bibitem[{Peng and Zhao(2009)}]{peng2009migration}
\bibinfo{author}{Peng, Z.}, \bibinfo{author}{Zhao, P.}, \bibinfo{year}{2009}.
\newblock \bibinfo{title}{Migration of early aftershocks following the 2004
  parkfield earthquake}.
\newblock \bibinfo{journal}{Nature Geoscience} \bibinfo{volume}{2},
  \bibinfo{pages}{877--881}.
%Type = Article
\bibitem[{Ross et~al.(2018)Ross, Meier, Hauksson and
  Heaton}]{ross2018generalized}
\bibinfo{author}{Ross, Z.E.}, \bibinfo{author}{Meier, M.A.},
  \bibinfo{author}{Hauksson, E.}, \bibinfo{author}{Heaton, T.H.},
  \bibinfo{year}{2018}.
\newblock \bibinfo{title}{Generalized seismic phase detection with deep
  learning}.
\newblock \bibinfo{journal}{Bulletin of the Seismological Society of America}
  \bibinfo{volume}{108}, \bibinfo{pages}{2894--2901}.
%Type = Article
\bibitem[{Ross et~al.(2017)Ross, Rollins, Cochran, Hauksson, Avouac and
  Ben-Zion}]{ross2017aftershocks}
\bibinfo{author}{Ross, Z.E.}, \bibinfo{author}{Rollins, C.},
  \bibinfo{author}{Cochran, E.S.}, \bibinfo{author}{Hauksson, E.},
  \bibinfo{author}{Avouac, J.P.}, \bibinfo{author}{Ben-Zion, Y.},
  \bibinfo{year}{2017}.
\newblock \bibinfo{title}{Aftershocks driven by afterslip and fluid pressure
  sweeping through a fault-fracture mesh}.
\newblock \bibinfo{journal}{Geophysical Research Letters} \bibinfo{volume}{44},
  \bibinfo{pages}{8260--8267}.
%Type = Article
\bibitem[{Saad and Chen(2020)}]{saad2020earthquake}
\bibinfo{author}{Saad, O.M.}, \bibinfo{author}{Chen, Y.}, \bibinfo{year}{2020}.
\newblock \bibinfo{title}{Earthquake detection and p-wave arrival time picking
  using capsule neural network}.
\newblock \bibinfo{journal}{IEEE Transactions on Geoscience and Remote Sensing}
  .
%Type = Article
\bibitem[{Saad and Chen(2021)}]{saad2021capsphase}
\bibinfo{author}{Saad, O.M.}, \bibinfo{author}{Chen, Y.}, \bibinfo{year}{2021}.
\newblock \bibinfo{title}{Capsphase: Capsule neural network for seismic phase
  classification and picking}.
\newblock \bibinfo{journal}{IEEE Transactions on Geoscience and Remote Sensing}
  .
%Type = Article
\bibitem[{Sha et~al.(2022)Sha, Faber, Gou, Liu, Li, Schramm, Stoecker,
  Steckenreiter, Vnucec, Wetzstein et~al.}]{sha2022multi}
\bibinfo{author}{Sha, Y.}, \bibinfo{author}{Faber, J.}, \bibinfo{author}{Gou,
  S.}, \bibinfo{author}{Liu, B.}, \bibinfo{author}{Li, W.},
  \bibinfo{author}{Schramm, S.}, \bibinfo{author}{Stoecker, H.},
  \bibinfo{author}{Steckenreiter, T.}, \bibinfo{author}{Vnucec, D.},
  \bibinfo{author}{Wetzstein, N.}, et~al., \bibinfo{year}{2022}.
\newblock \bibinfo{title}{A multi-task learning for cavitation detection and
  cavitation intensity recognition of valve acoustic signals}.
\newblock \bibinfo{journal}{arXiv preprint arXiv:2203.01118} .
%Type = Article
\bibitem[{Sleeman and Van~Eck(1999)}]{sleeman1999robust}
\bibinfo{author}{Sleeman, R.}, \bibinfo{author}{Van~Eck, T.},
  \bibinfo{year}{1999}.
\newblock \bibinfo{title}{Robust automatic p-phase picking: an on-line
  implementation in the analysis of broadband seismogram recordings}.
\newblock \bibinfo{journal}{Physics of the earth and planetary interiors}
  \bibinfo{volume}{113}, \bibinfo{pages}{265--275}.
%Type = Incollection
\bibitem[{St-Onge(2011)}]{st2011akaike}
\bibinfo{author}{St-Onge, A.}, \bibinfo{year}{2011}.
\newblock \bibinfo{title}{Akaike information criterion applied to detecting
  first arrival times on microseismic data}, in: \bibinfo{booktitle}{SEG
  Technical Program Expanded Abstracts 2011}. \bibinfo{publisher}{Society of
  Exploration Geophysicists}, pp. \bibinfo{pages}{1658--1662}.
%Type = Article
\bibitem[{Stehman(1997)}]{stehman1997selecting}
\bibinfo{author}{Stehman, S.V.}, \bibinfo{year}{1997}.
\newblock \bibinfo{title}{Selecting and interpreting measures of thematic
  classification accuracy}.
\newblock \bibinfo{journal}{Remote sensing of Environment}
  \bibinfo{volume}{62}, \bibinfo{pages}{77--89}.

\end{thebibliography}

\end{document}